\documentclass[a4paper,
               keeplastbox,   
               ]{jacow}
\usepackage{pdfpages,multirow,ragged2e} %
\usepackage[Symbol]{upgreek}
\usepackage{color}  
\usepackage{hyperref}
\usepackage{graphics}
\usepackage{subcaption}
\hypersetup{
    colorlinks=false,       
    linkbordercolor=green,          
    citecolor=black,        
    filecolor=magenta,      
    urlcolor=blue           
}

\makeatletter%
	\ifboolexpr{bool{xetex}}
	 {\renewcommand{\Gin@extensions}{.pdf,%
	                    .png,.jpg,.bmp,.pict,.tif,.psd,.mac,.sga,.tga,.gif,%
	                    .eps,.ps,%
	                    }}{}
\makeatother

\ifboolexpr{bool{xetex} or bool{luatex}} 
 {}                                      
 {\usepackage[utf8]{inputenc}}           

\usepackage[USenglish]{babel}

\ifboolexpr{bool{jacowbiblatex}}%
 {%
  \addbibresource{jacow-test.bib}
  \addbibresource{biblatex-examples.bib}
 }{}
\begin{document}

\newcommand{\pt}{\ensuremath{p_\mathrm{T}}}
\newcommand{\pT}{\pt}
\newcommand{\pp}{\ensuremath{\mbox{p}\mbox{p}~}}
\newcommand{\Minv}{\ensuremath{M_\mathrm{inv}}}
\newcommand{\Qinv}{\ensuremath{Q_\mathrm{inv}}}

\title{Photon pair distributions and event-mixing at low invariant mass}

\title{Photon pair distributions and event-mixing at low invariant mass}

\author{M. Sas\textsuperscript{1,2}, L. Lamers\textsuperscript{2}, T. Peitzmann\textsuperscript{2}\\
        \textsuperscript{1}Physics Department, Yale University, New Haven CT, U.S.A.\\
		\textsuperscript{2}Institute for Subatomic Physics, Utrecht University/Nikhef, Utrecht, Netherlands\\
		\today}
	
\maketitle

\begin{abstract}
Neutral meson cross-section and direct photon HBT measurements in high energy pp, pA, and AA collisions rely on either subtracting or dividing out the background underlying the $\Minv$ or $\Qinv$ distributions.
   In this paper we investigate the photon pair distributions for simulated proton-proton collisions at $\sqrt{s}=13~$TeV, and study why the background in $\Minv$ gets poorly described by the conventional event-mixing method implemented in experimental analyses. We show that it fails to describe the background as it gives a qualitatively different shape especially at invariant masses around the $\pi^{0}$ and $\eta$ meson mass, which is mainly caused by the correlations originating from parton fragmentation. 
\end{abstract}

\section{Introduction}
\begin{figure}[!htb]
   \centering
    \includegraphics[width=0.99\columnwidth]{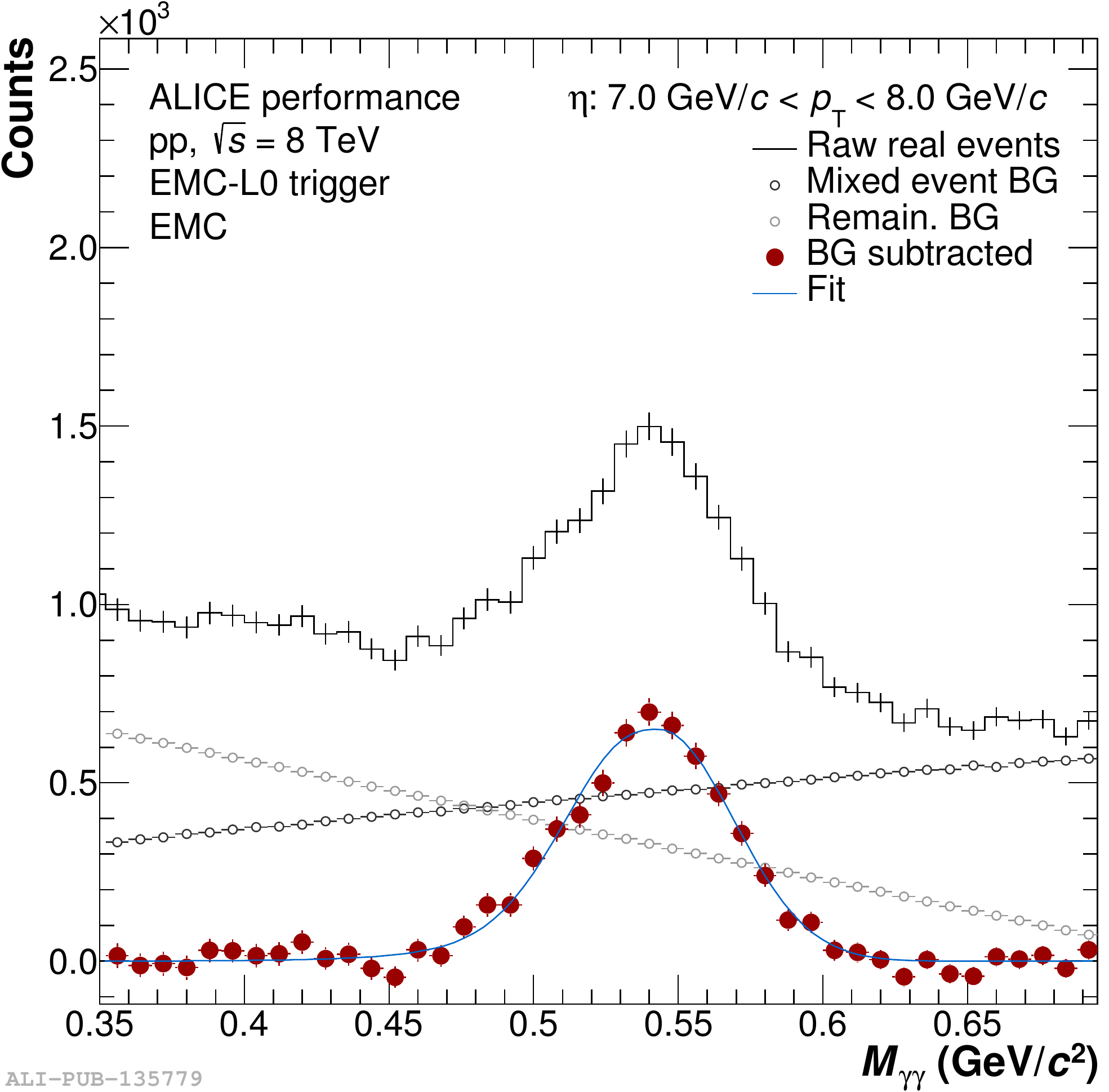}
   \caption{The diphoton $\Minv$ distribution around the $\eta$ meson mass for $7.0<p_{\mathrm{T,pair}}<8.0$ GeV/c, as measured with the ALICE EMCal detector in pp collisions at $\sqrt{s}=8~$TeV. It shows that the background is not described by the distribution obtained from event-mixing.}
   \label{fig:ALICE_Eta}
\end{figure}

Pair correlation analysis is a standard analysis technique in particle and nuclear physics. One of the main examples is the study of two-body decays via the invariant mass spectrum of potential decay daughters, where one can extract the yield of the mother particles from a correlated peak in the pair distribution atop a broader background from random pairs, the so-called combinatorial background \cite{NeutrMesPaper3, NeutrMesPaper1, NeutrMesPaper2, NeutrMesPaper4, ALICElightflavor, ALICEheavyflavor}. Another example is that of femtoscopy \cite{femto2,femto3}, the study of (most often identical) pair correlations in momentum space, where quantum statistics and/or interactions cause a correlation, often at very small relative momentum. In all of these cases one has to find a means to describe the background. Sometimes, e.g. in invariant mass analysis, the correlation produces a prominent, distinct structure on a very smooth background, which makes the description of the latter possible with simple analytical functions. There are, however, many cases where high precision measurements require  knowledge about the shape of the background. This is important, when the phase space acceptance of the detector leads to a non-monotonous shape of the background near the signal. A widely used method to estimate the shape of the combinatorial background is that of event mixing, i.e. combining particle pairs from different events, which by construction will not be correlated.
Below, we will use the example of photon-pair distributions, but our analysis should be relevant also for the analysis of other pair-correlations.\\
The yield of light neutral mesons, such as the $\pi^{0}$ and $\eta$ meson, is calculated by integrating the signal peak in the diphoton invariant mass distribution on top of a combinatorial background. The invariant mass distribution can either be fitted simultaneously with a Gaussian describing the signal peak and a polynomial describing the background, or the background can first be subtracted using the event-mixing method. Also, for direct photon HBT measurements the correlator $C(Q_{\mathrm{inv}})=A(Q_{\mathrm{inv}}) / B(Q_{\mathrm{inv}})$ divides the same-event pair distribution by the event-mixing distribution. For such measurements it is important that the combinatorial background which is present in same-event photon pairs is properly described by the distributions obtained with the event-mixing method.\\
However, as demonstrated through an example shown in Fig.~\ref{fig:ALICE_Eta}, the combinatorial background obtained with the event-mixing method fails to describe the background of the same-event pair distribution \cite{NeutrMesPaper3}. In this case the residual background is subtracted via a linear fit, which only works well if the background underneath the peak is well behaved. In fact, a qualitatively similar correlation effect has been observed in femtoscopic analysis (see e.g. \cite{femto2}).\\
In this paper we investigate the photon pair distributions ($0<\Minv<5~$GeV/$c^{2}$), for three different pair \pT~ intervals below $10~$GeV/$c$ in simulated proton-proton collisions at $\sqrt{s}=13~$TeV, using the PYTHIA$8.2$ \cite{pythia} event generator, and discuss the performance of describing the background using the event-mixing method also in more critical cases.\\
The article is divided into the following sections: In Sec.~2 we define the event-mixing method used to obtain the results. In Sec.~3 we present the diphoton invariant mass distributions and the estimated background using the event-mixing method. In Sec.~4 we discuss the performance of event-mixing when applying additional detector constraints. In Sec.~5 we investigate the origin of the correlated background. In Sec~6. we attempt to improve the conventional way of event-mixing, i.e. to also capture the correlated part of the combinatorial background. In Sec~7. we conclude with a summary.

\section{Event-mixing method}

Using the event-mixing method, the combinatorial background is estimated by combining photons exclusively from different events. These photon pairs never share a common ancestor and thus make sure that there are no resonances. The method can also describe non-trivial shapes of the combinatorial background in the signal regions, which for a polynomial fit would be impossible. These non-trivial background shapes are usually created by detector constraints and non-homogeneous acceptance, such as calorimeter module sizes or dead calorimeter cells/modules. As can be seen in \cite{NeutrMesPaper3, NeutrMesPaper1, NeutrMesPaper2, NeutrMesPaper4}, this remains to be a challenge for especially the $\eta$ meson.\\
For the results presented in this paper we pair all current event photons with all photons from all events in the mixing-pool. The mixing-pool acts as a First-In-First-Out (FIFO) buffer storing the last $200$ events. After the current event is processed the event is added to the mixing-pool if it has at least two photons. Also, it is customary to define multiple mixing-pools, each having distinct event characteristics, such as charged/neutral particle multiplicity and z-vertex location. The photons from the current event are then paired with the events from the pool which shares the same event characteristics. The z-vertex location is only important to take into account for experimental data, since PYTHIA has a fixed z-vertex. In addition, we have found that mixing events with within a similar multiplicity interval is not changing the distribution, as long as the buffer is large and the single particle energy spectra are the same. The latter would not be the case for heavy-ion collisions, where the nuclear matter effects have shown to radically change the shapes for different collision centralities. For this study it suffices to use a single mixing-pool of $200$ events.

\section{Diphoton invariant mass distributions}

The results presented in this paper are based on PYTHIA-generated \cite{pythia} proton-proton collisions at $\sqrt{s}=13~$TeV, using two data-sets; SoftQCD, and HardQCD with a minimum $\hat{p}_{\mathrm{T}}$ of $20~$GeV/$c$. This enables us to study the diphoton invariant mass distributions in minimum bias (MB) and high-\pT~ jet type events. In the data-set with HardQCD, the minimum $\hat{p}_{\mathrm{T}}$ of $20~$GeV/$c$ ensures that we have high-\pT~ partons, enhancing the correlated part of the pair distributions. The diphoton invariant mass is calculated as

\begin{align*}
    \Minv &= \sqrt{2 E_{\gamma,1} E_{\gamma,2} (1- \cos \psi)}\\
          &= (E_{\gamma,1} + E_{\gamma,2}) \sqrt{\frac{1-\alpha^2}{2} (1 - \cos \psi)}
\end{align*}

where $E_{\gamma}$ is the energy of the photon, $\psi$ is the opening angle between the two photons, and $\alpha$ is the energy asymmetry with $\alpha = (E_{\gamma,1}-E_{\gamma,2})/(E_{\gamma,1}+E_{\gamma,2})$. For same-event pair combinations, the $\pi^{0}$ and $\eta$ will create distinct peaks at $\Minv=134.98~$GeV/$c^{2}$ and $\Minv=547.86~$GeV/$c^{2}$, respectively. Due to physics effects such as jet production, the energy and opening angle distributions from the two combined photons are not necessarily qualitatively the same for same-event and event-mixing pair distributions. This will then be apparent from the comparison of the $\Minv$ distributions.

\begin{figure}[!htb]
   \centering
    \includegraphics[width=\columnwidth]{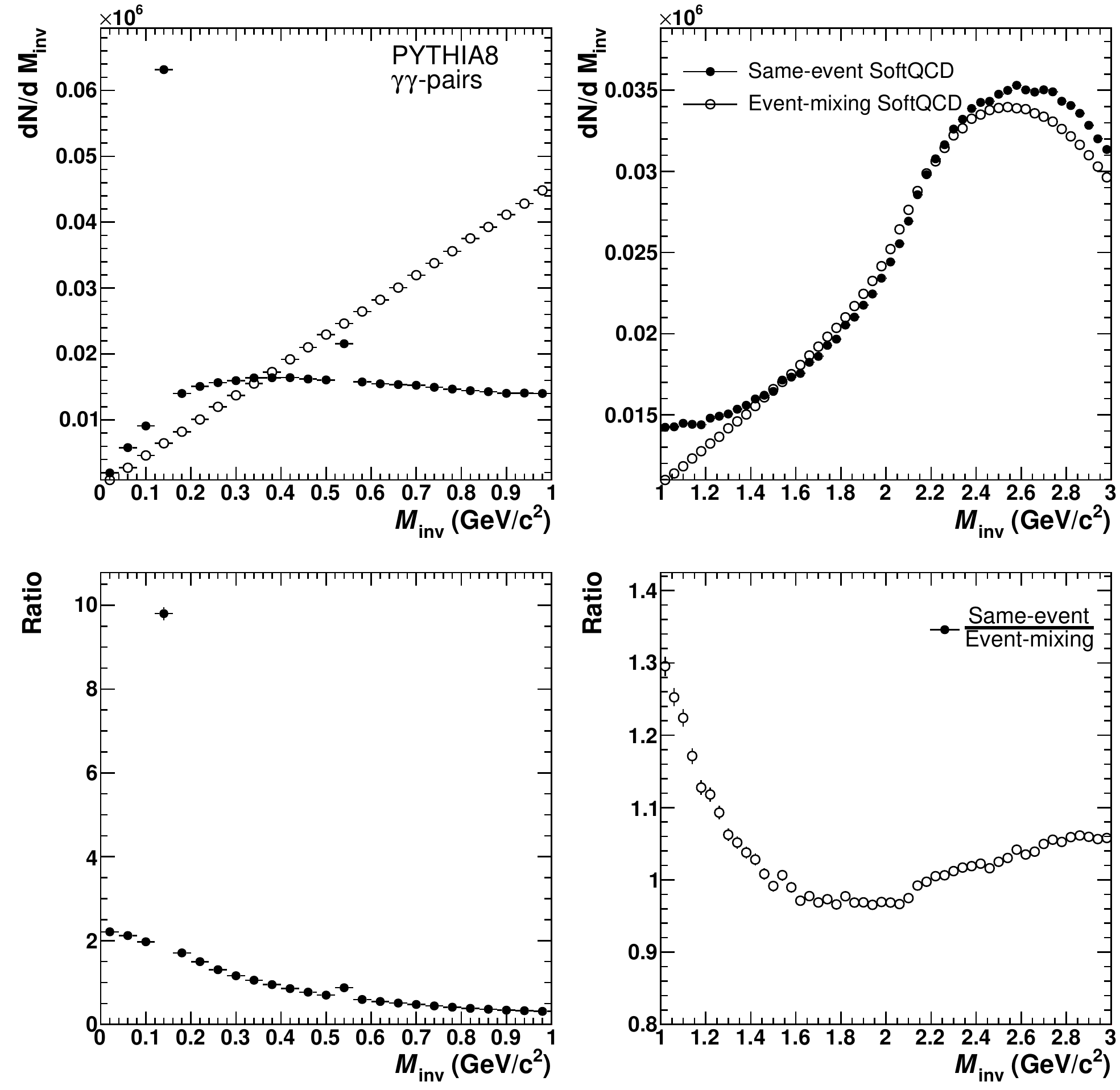}
   \caption{The diphoton invariant mass distribution obtained from PYTHIA (SoftQCD), for $0<\Minv<1.0$ (top left) and $1.0<\Minv<3~$GeV/$c^{2}$ (top right) for same event pairs (closed) and event-mixing (open), for a diphoton pair momentum of $1<\pT<2~$GeV/$c$. The bottom panels show a ratio of the same-event pair to the event-mixing distributions.}
   \label{fig:InvMassZoomed_hard}
\end{figure}

In Fig.~\ref{fig:InvMassZoomed_hard} we show the diphoton invariant mass distribution obtained from PYTHIA (SoftQCD) for $0<\Minv<1.0~$GeV/$c^{2}$ and $1.0<\Minv<3.0~$GeV/$c^{2}$, for same-event pairs (filled markers) and for event-mixing (open markers). The event-mixing distribution is scaled to match the same-event pair distributions, at $0.35~$GeV/$c^{2}$ for the upper left panel and at $1.5~$GeV/$c^{2}$ for the upper right panel. The figure indicates that event mixing is unable to describe the shape of the $\Minv$ distribution around the $\pi^{0}$ and $\eta$ mass, and more generally is unable to capture the background of photon pairs for $0<\Minv<3.0~$GeV/$c^{2}$. For $\Minv<0.2~$GeV/$c^{2}$, the same-event pairs show a much steeper distribution compared to the event-mixing distribution. Furthermore, the shape of the same-event pairs flattens for $\Minv>0.3~$GeV/$c^{2}$, while the event-mixing distribution continues to rise. The simulated PYTHIA events thus reproduce the observations made from Fig.~\ref{fig:ALICE_Eta}.


\begin{figure*}[htb!]
    \centering
    \includegraphics[width=0.93\textwidth]{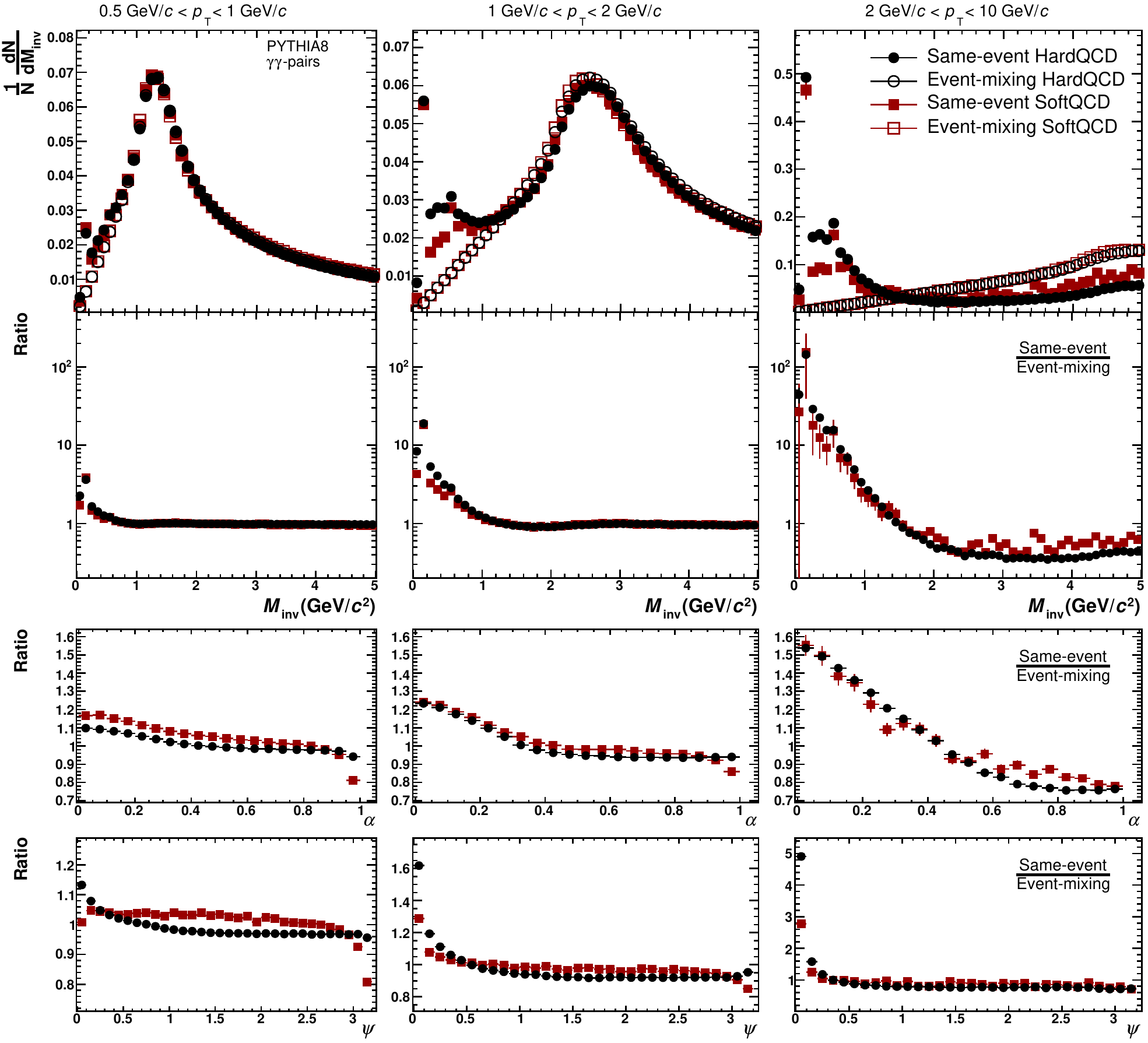}
    \caption{The diphoton $\Minv$ distributions for PYTHIA HardQCD (black) and SoftQCD (red), for same-event pair distributions (filled) and event-mixing (open) symbols. The lower panels show the ratios of same-event pairs to event-mixing pairs, for the $\Minv$, energy asymmetry $\alpha = (E_{\gamma,1}-E_{\gamma,2})/(E_{\gamma,1}+E_{\gamma,2})$, and opening angle $\psi$ distributions.}
    \label{fig:Combination_hard_soft}
\end{figure*}

Next, we compare the PYTHIA HardQCD to PYTHIA SoftQCD. In SoftQCD the particle production is not dominated by high-$\pT$~ parton fragmentation as with HardQCD, but rather by the soft component or underlying event. The results are shown in Fig.~\ref{fig:Combination_hard_soft}. There is a dominant correlation effect for $\Minv<1.0~$GeV/$c^{2}$, which shows that event-mixing does not describe the same-event pair distributions. The effect also increases with pair-$\pT$, indicating that this is related to jet fragmentation. For $\Minv>2.0~$GeV/$c^{2}$, the shapes of the $\Minv$ distributions from same-event and event-mixing photon pairs for both SoftQCD and HardQCD events match within about $\pm 2\%$, $\pm 5\%$, and $\pm 10\%$ for pair-$\pT$ intervals of $0.5 < \pT < 1$ GeV/$c$, $1 < \pT < 2$ GeV/$c$, and $2 < \pT < 10$ GeV/$c$, respectively, and rapidly start to diverge for lower $\Minv$ (also see Appendix Fig.~\ref{fig:Combination_hard_soft_zoom}). As can be seen in the ratio of the $\Minv$ distributions, the same-event and event-mixing distributions of SoftQCD events are more in agreement compared to distributions of the HardQCD events. This is probably due to the decrease in correlated background in the same-event pair distributions. Furthermore, the event-mixing pair distribution for HardQCD and SoftQCD production is within uncertainties the same, showing that they both reproduce the combinatorial background to the same extent, and are both unable to describe the same-event pair distributions especially at lower $\Minv$. The explicit dependence of the energy asymmetry $\alpha = (E_{\gamma,1}-E_{\gamma,2})/(E_{\gamma,1}+E_{\gamma,2})$ and opening angle $\psi$ on $\Minv$ show a deviation that increases for lower values of $\alpha$ and $\psi$ in all $\pT$ bins.


\section{Pseudorapidity and azimuthal angle restrictions}

\begin{figure*}[htb]
    \centering
    \includegraphics[width=0.93\textwidth]{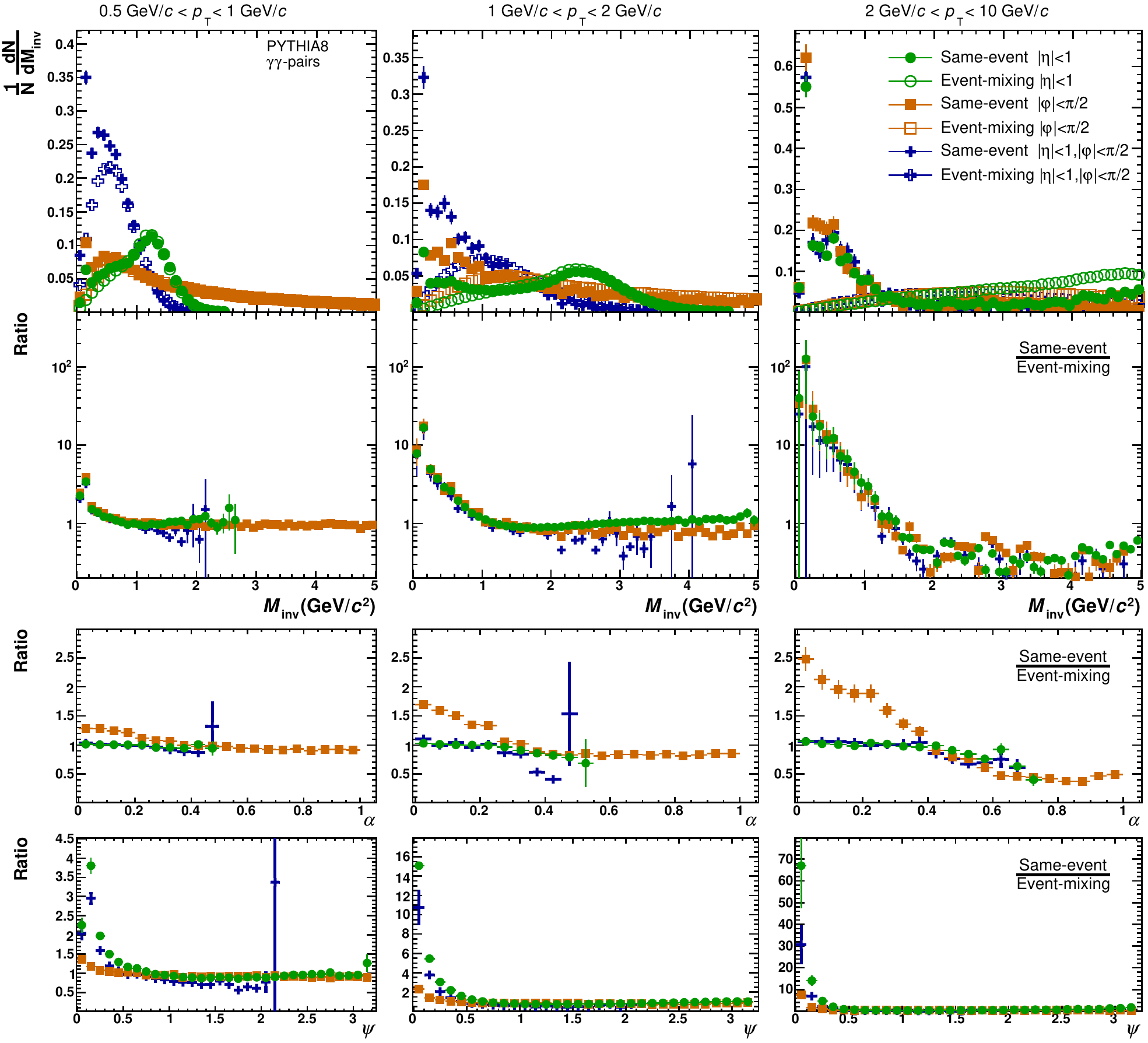}
    \caption{Diphoton $\Minv$ distributions for same-event and event-mixing pairs, for the detector constraints; $|\eta| < 1$, $\varphi < \pi/2$, and $|\eta| < 1$ \& $\varphi < \pi/2$. The lower panels show the ratios of same-event pairs to the event-mixing pair distributions, for the $\Minv$, energy asymmetry $\alpha = (E_{\gamma,1}-E_{\gamma,2})/(E_{\gamma,1}+E_{\gamma,2})$, and opening angle $\psi$ distributions.}
    \label{fig:EtaPhiRestrictions}
\end{figure*}

In experiments we have to deal with the actual size of the detector, and this constrains the pseudorapidity $\eta$ and/or the azimuthal angle $\varphi$ regions from which we can obtain the photons and introduces a purely kinematic effect. This limits the opening angle of the diphoton pair and changes the same-event and  $\Minv$ distributions in a non-trivial way, while the distributions from event-mixing do change in a trivial way. Furthermore, since the average opening angle of a $\pi^{0}$ decay is $\pT$~ dependent, detector constraints will affect low and high diphoton $\pT$~ distributions differently.\\
In Fig.~\ref{fig:EtaPhiRestrictions}, using the HardQCD dataset, we introduce the following detector constraints; $|\eta|<1$, $\varphi<\pi/2$, and the combination of the two. The results for restricting the photons in $\eta$ show a large improvement in the $\alpha$ distributions. Jet correlations show a constraint in $\Delta \eta$, which translates into a constraint in $\alpha$ for correlated pairs, when $\pT$ is fixed. The true pairs are thus found more closely together in $\alpha$ compared to pairs from event mixing, which can be more widespread in $\eta$. In this way, an apparent correlation in the energy difference shows up, even though the physical mechanism may introduce only a correlation in angle. The difference disappears, if an artificial constraint in $\eta$ is introduced for all pairs, and in this case the correlation as a function of $\alpha$ also disappears. The restriction in $\varphi$ shows similar results compared to the unrestricted case, because a difference in $\varphi$ does not change the relation between $\pT$ and $E$, opposite to a difference in $\eta$. Detector constraints do change the distributions, but the combinatorial background at low $\Minv$ is still poorly described by the event-mixing method.

\section{Diphoton pairs from same and different ancestors}

\begin{figure*}[htb]
    \centering
    \includegraphics[width=0.93\textwidth]{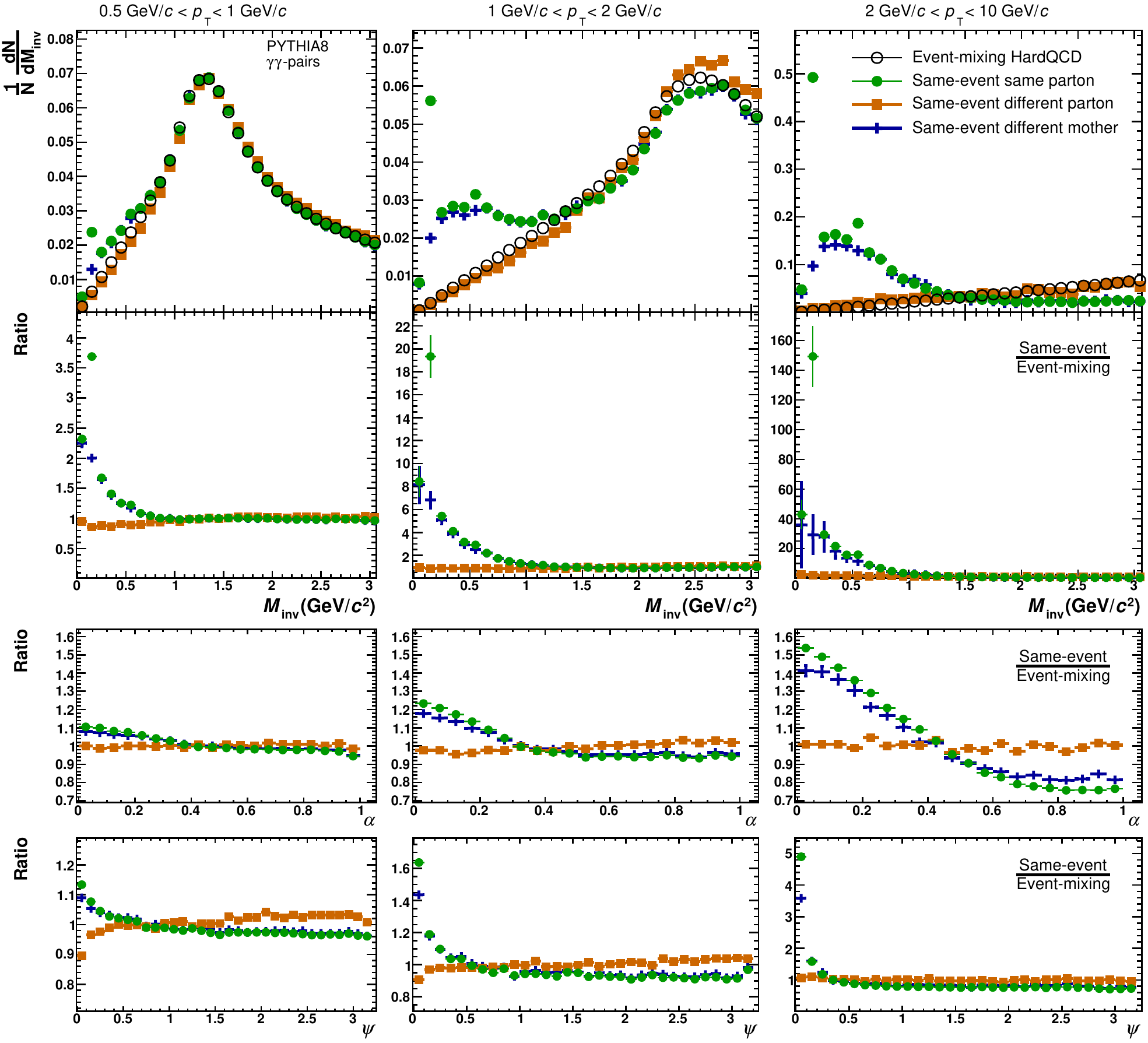}
    \caption{The diphoton $\Minv$ mass and the ratios for same-event pairs and event mixing for the following three cases; 1. when the photon pair for same-event-mixing shares a common parton ancestor, 2. when both photons do not share a common parton ancestor, and 3. when the photon pair do not share the same mother- or grandmother particle whenever it considers a hadronic decay. In each case the ratio is taken to the same event-mixing pair distributions.
    }
    \label{fig:PartonMixing}
\end{figure*}

The initial hard scatterings simulated by PYTHIA produce partons that fragment and give rise to final state hadrons. These partons are the most dominant source for particle production and all final state particles sharing the parton as common ancestor will thus have correlations in $\Delta \eta$ and $\Delta \varphi$. To investigate whether this affects the diphoton mass for same-event pairs and event-mixing pairs, using the HardQCD dataset, we study the following three cases; 1. when the photon pair shares a common parton ancestor, 2. when both photons do not share a common parton ancestor, and 3. when the photon pair does not originate from one hadronic decay, such as the $\pi^{0}$ or $\eta$. In this last case the photons can still share a common (parton) ancestor. For all the three cases the event-mixing procedure is the same, namely all photons come from different events and thus ancestors.\\ The results for the three different cases are shown in Fig.~\ref{fig:PartonMixing}. First, photon pairs coming from the same parton give qualitatively the same results compared to the earlier distributions. When we compare this to case 3, we see that except for the $\pi^{0}$ and $\eta$ peak being removed, there is only a small difference that we attribute partly to the three-pion decay of the $\eta$ meson. For case 2, where both photons do not share a common parton ancestor, it is shown that the event-mixing is now describing the whole invariant mass distribution much better also at lower $\Minv$ (see also Appendix Fig.~\ref{fig:PartonMixing_zoom}). Both the observations at lower and higher $\Minv$ are now shown to be related to the jet fragmentation, where the effects at lower and higher $\Minv$ are then related to the near- and away-side jet structures, respectively. A qualitatively similar effect at low $\Minv$ was already seen in studies at RHIC \cite{STARpi0}, but the jet correlations observed were significantly weaker than what we find here at LHC energy. The ratios of the energy asymmetry $\alpha$ and opening angle $\psi$ are also individually much closer to unity. This indicates that the inability of the event-mixing method to properly describe the combinatorial background is due to the fact that the method does not conserve the particle to particle correlations of the parton shower. It should be noted, that femtoscopy measurements also show a residual pair correlation at small momentum difference, which is in that case attributed to mini-jets \cite{femto2}.

\section{Modified diphoton event-mixing}

\begin{figure*}[htb!]
    \centering
    \includegraphics[width=0.93\textwidth]{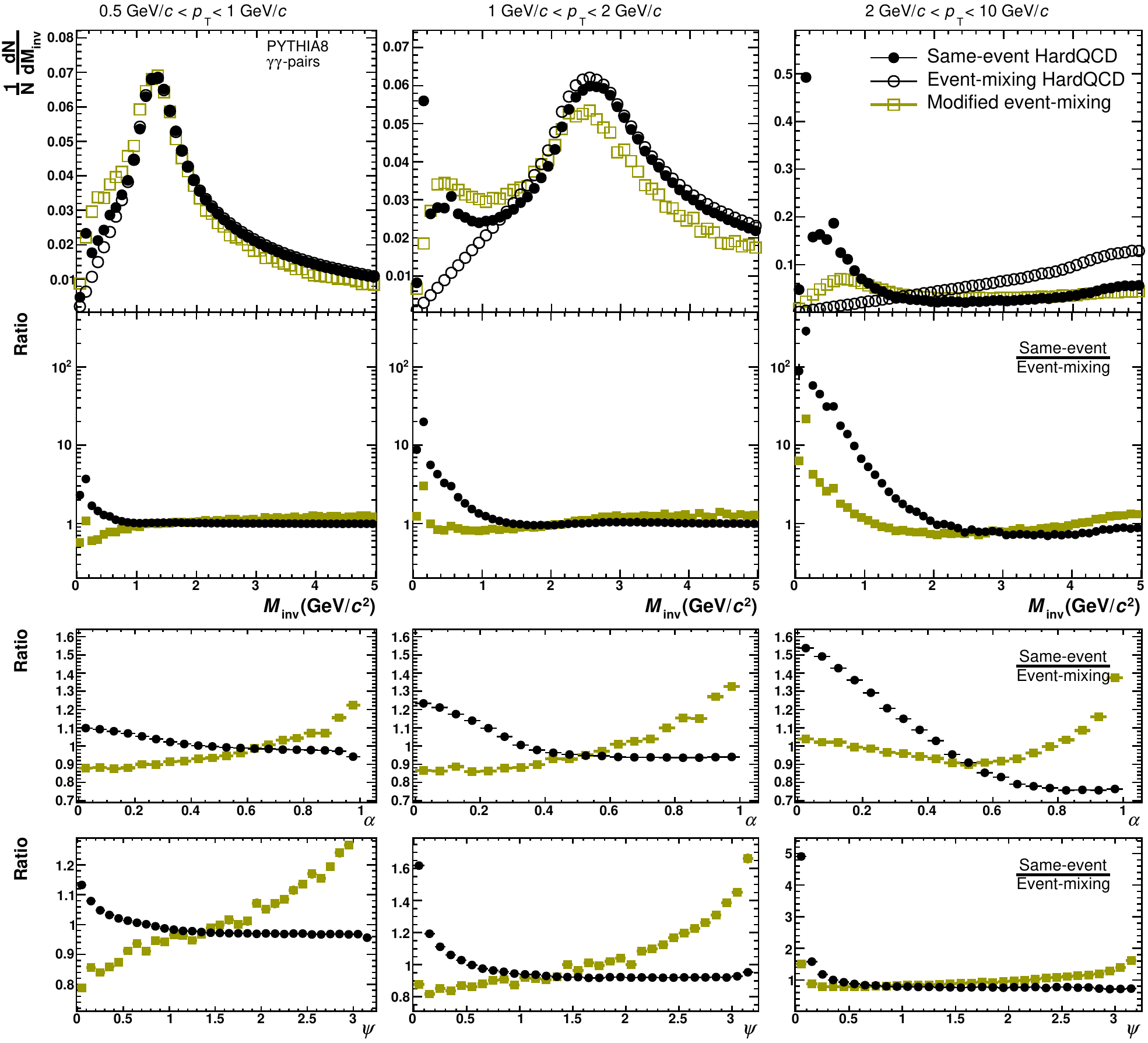}
    \caption{The diphoton $\Minv$ distributions for same-event pair distributions (filled circle), event-mixing (open circle), and modified event-mixing (open square). For the modified event-mixing, photon are only paired when the parton ancestor is within $|\eta|<0.25$ and $|\varphi|<0.25$. The lower panels show the ratios of same-event pairs to event-mixing pairs, for the $\Minv$, energy asymmetry $\alpha = (E_{\gamma,1}-E_{\gamma,2})/(E_{\gamma,1}+E_{\gamma,2})$, and opening angle $\psi$ distributions.}
    \label{fig:Modified-event-mixing}
\end{figure*}

The previous section showed that pairing photons from the same parton, even when the photons do not come from the same hadronic decay, creates a broad correlated background in the region of the $\pi^{0}$ and $\eta$ meson mass. Pairing photons that originate from different partons do destroy this correlation, but would make the neutral meson measurement practically impossible. In order to attempt to improve the event-mixing, such that it describes the correlated background in the low $\Minv$ region, we can choose to only pair photons if the parton ancestor from both photons are close in $\eta$ and $\varphi$. This can partly preserve the $\alpha$ and $\psi$ distributions of the same-event pairs with respect to the event-mixing pairs. We chose to only pair the photons if the parton ancestor is within $|\eta|<0.25$ and $|\varphi|<0.25$, and  ensures that both parton ancestors are pointing in the same direction while also allowing for sufficient combinatorics to be generated. An even smaller selection window does not significantly improve the results, as the two parton showers also vary in final state multiplicity and angles between the constituents. The result in Fig.~\ref{fig:Modified-event-mixing} shows that the $\Minv$ distribution from the modified event-mixing has a qualitatively different shape and describes the same-event pair distributions better. However, for the first two $\pT$~ intervals, the modified event-mixing overestimates the background in real events, while for $2.0<\pT<10~$GeV/$c$, it underestimates the background. The behaviour at low $\pT$ could be due to the fact that not all photons in that case come from a single jet, so the restriction on the particle from the direction of the parton in the event mixing is stronger than in real data. On the other hand, at higher $\pT$ the correlation may be stronger in the same-event pair distributions, because the uncertainty of the parton direction in $\eta$ and $\phi$ is not negligible compared to the width of a jet, and thus leads to a dilution of the correlation in mixed events.\\

\section{CONCLUSION}

In this paper the diphoton invariant mass distributions are investigated for simulated PYTHIA HardQCD and SoftQCD events. We have shown that for both lower and higher $\Minv$, the same-event pair distributions are not fully described by the distributions obtained by the event-mixing technique. For lower invariant masses typically\\ around the $\pi^{0}$ and $\eta$ meson mass, as shown in Fig.~\ref{fig:Combination_hard_soft}, the event-mixing method completely fails to describe the distribution as it gives a qualitatively different shape. In addition, detector constraints do change the $\Minv$ distributions, but the same-event pairs are similarly (un)described by the event-mixing, as can be seen in Fig.~\ref{fig:EtaPhiRestrictions}.\\
In Fig.~\ref{fig:PartonMixing}, we studied three cases; 1. when the photon pair shares a common parton ancestor, 2. when both photons do not share a common parton ancestor, and 3. when the photon pair do not share the same mother- or grandmother particle whenever it considers an hadronic decay. From this we conclude that the correlations at lower $\Minv$ is not caused by decay kinematics, but rather by the parton fragmentation into final state hadrons.\\
In Fig.~\ref{fig:Modified-event-mixing}, we attempted to improve the quality of event-mixing by pairing photons only when the parton ancestor is within $|\eta|<0.25$ and $|\varphi|<0.25$, preserving the $\alpha$ and $\psi$ distributions of the same-event pairs with respect to the event-mixing pairs. This improves the description of the combinatorial background by a factor of 2-10 in the region of the $\pi^{0}$ and $\eta$ meson mass, while it doesn't improve the description for higher masses. Experimentally this approach is more difficult to implement, but jet-finding algorithms with a low minimum jet $\pT$~ can approximate the direction of the parton that produces the photon. In \cite{STARpi0} the qualitatively similar residual correlations were described with a simple jet-aligned mixing, however for the much stronger effect we see, a similarly simple approach was unfortunately not successful. Therefore, we argue that it might be best to only use the event-mixing technique to subtract the uncorrelated background, as this is the only background that can be calculated reliably without introducing further biases.\\
These observations have implications for the ability to describe the correlated and uncorrelated part of the diphoton invariant mass background using the event-mixing method, which is used for neutral meson as well as for direct photon HBT measurements. Both these measurements rely on either subtracting or dividing out the background from wrong-pair combinations, and it would be interesting to see whether jet-aligned mixing can improve the background description also for experiments.

\section{ACKNOWLEDGEMENTS}
This work was supported in part by the Netherlands Organisation for Scientific Research (NWO) and the Office of Nuclear Physics of the U.S. Department of Energy. Work supported by the US DOE under award number DE-SC004168.

%
%
\ifboolexpr{bool{jacowbiblatex}}%
	{\printbibliography}%
	{%
	

}

\clearpage
\onecolumn
\appendix
\section{APPENDIX}

\begin{figure}[htb]
    \centering
    \includegraphics[width=0.9\textwidth,trim={0 7.38cm 0 0},clip]{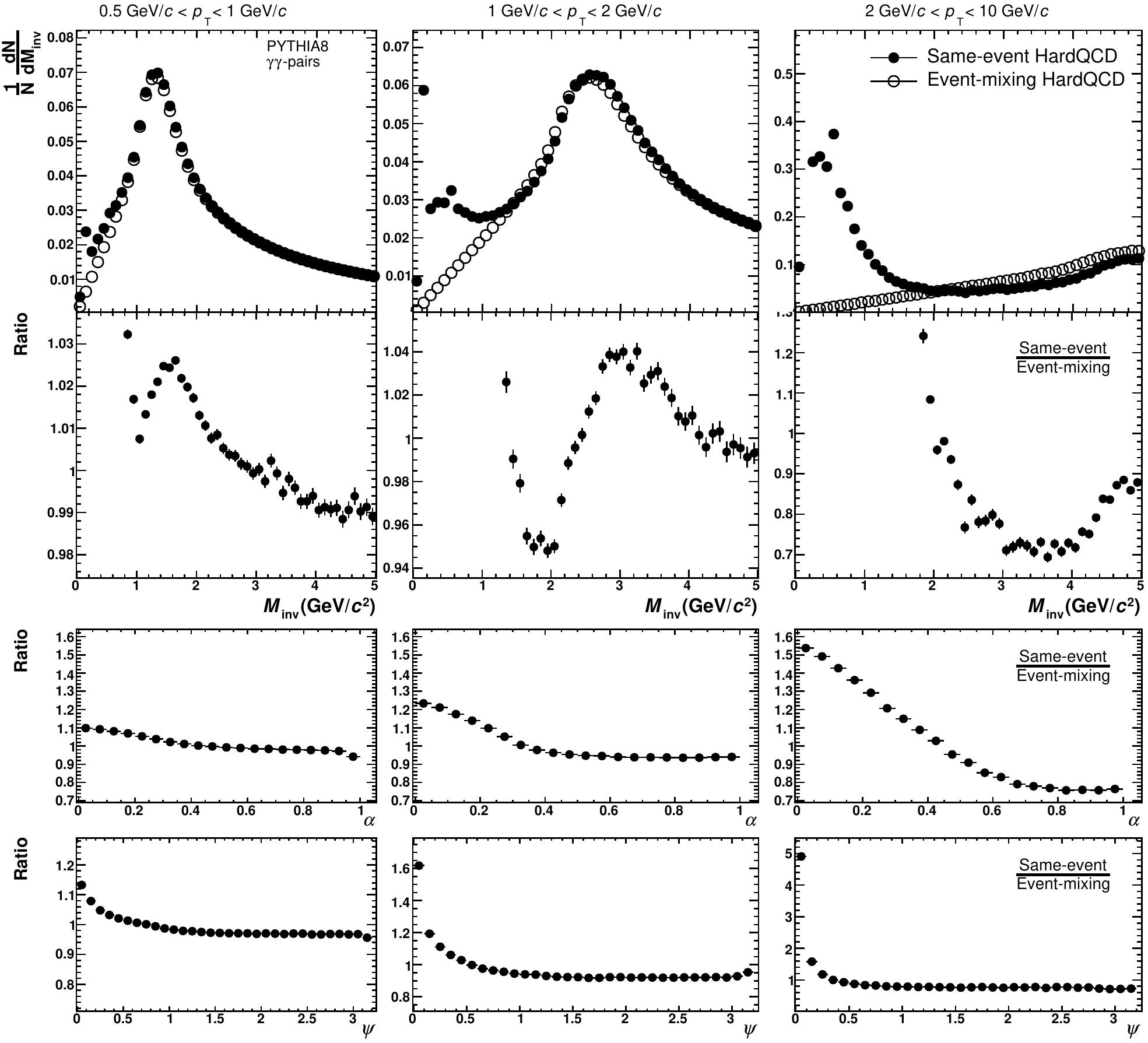}
    \caption{The diphoton $\Minv$ distributions for PYTHIA8 HardQCD, for same-event pair distributions (filled) and event-mixing (open) symbols. The lower panels show the ratios of same-event pairs to event-mixing pairs over a wide mass range. The distributions from the PYTHIA8 SoftQCD dataset has been omitted as it doesn't have the appropriate statistical precision.}
    \label{fig:Combination_hard_soft_zoom}
\end{figure}

\begin{figure}[htb]
    \centering
    \includegraphics[width=0.9\textwidth,trim={0 7.38cm 0 0},clip]{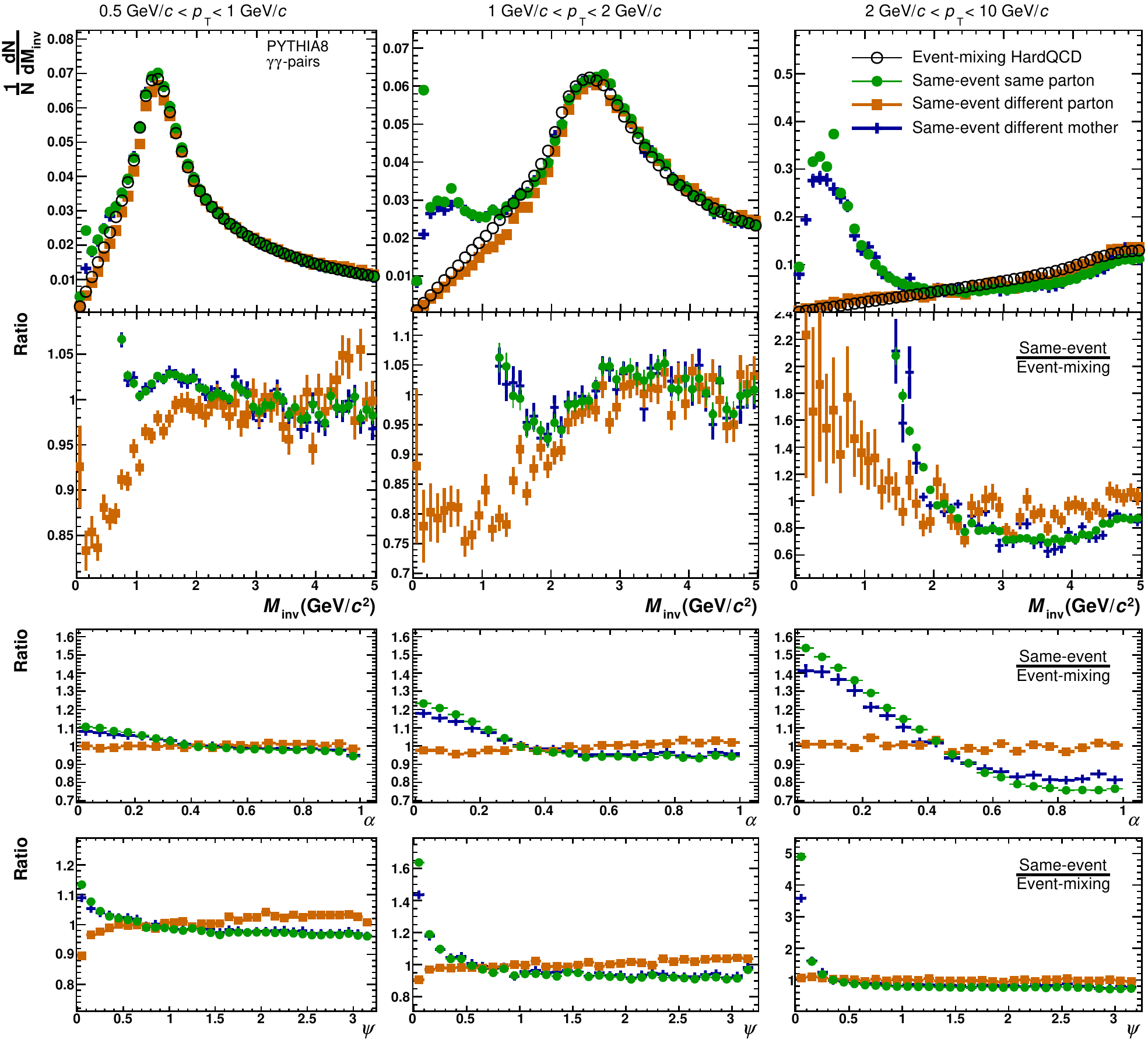}
    \caption{The diphoton $\Minv$ mass and the ratios for same-event pairs and event mixing over a wide mass range for the following three cases; 1. when the photon pair for same-event-mixing shares a common parton ancestor, 2. when both photons do not share a common parton ancestor, and 3. when the photon pair do not share the same mother- or grandmother particle whenever it considers a hadronic decay.
    }
    \label{fig:PartonMixing_zoom}
\end{figure}

The paper discusses mainly the behaviour of the relatively strong correlation seen in the continuous background of the two-photon invariant mass spectra at low masses $(\Minv<2~$GeV/$c^{2})$, which leads to a discrepancy between the background in real data and the event mixing. We had noted there, that the background description is reasonably good for larger masses. However, even at larger masses discrepancies are observed, albeit on a much smaller level, which we want to briefly discuss here, even when they are not relevant for the main analysis in the paper. In Fig.~\ref{fig:Combination_hard_soft_zoom} and Fig.~\ref{fig:PartonMixing_zoom} the ratio panels are shown such that the discrepancies at higher masses are visible. Figure \ref{fig:Combination_hard_soft_zoom} indicates that the shapes of the $\Minv$ distributions from same-event and event-mixing photon pairs for both SoftQCD and HardQCD events match within $\pm 2\%$, $\pm 5\%$, and $\pm 10\%$ for pair-$\pT$ intervals of $0.5 < \pT < 1~$GeV/$c$, $1 < \pT < 2~$GeV/$c$, and $2 < \pT < 10~$GeV/$c$, respectively. However, within these limits one observes interesting characteristic structures: In particular for the two lower $\pT$ bins there is a minimum in the ratio at intermediate mass, and a maximum at $\Minv \approx 1.5$ and $\approx 3$ GeV/$c$, respectively. Most likely, this structure is due to away-side jet-like correlations, as the mass values are close to twice the average energy of the two photons used. For the highest $\pT$ bin, a similar maximum would be outside of the mass range and thus not be visible. Figure \ref{fig:PartonMixing_zoom} shows that for higher masses all the different cases are described similarly well by the event-mixing method. The main feature here, as described in the main text, is the fact that the distribution obtained by pairing photons that originate from a different parton do not have the large enhancement found at lower $\Minv$. In the ratio plot it shows that the same-event distribution decreases significantly for lower $\Minv$ for the first two $\pT$-intervals. The combination of the two cases, the pairs from the same parton, which show the strong enhancement at low mass, and the pairs from different partons, which show a depletion, most likely leads to the characteristic structure in Fig.~\ref{fig:Combination_hard_soft_zoom}. A further study of this additional correlation phenomenon is beyond the scope of the paper.

%
%


\end{document}